CrossMark

# Classical Simulated Annealing Using Quantum Analogues

**Brian R. La Cour[1]** · **James E. Troupe[1]** · **Hans M. Mark[1]**

**Abstract** In this paper we consider the use of certain classical analogues to quantum tunneling behavior to improve the performance of simulated annealing on a discrete spin system of the general Ising form. Specifically, we consider the use of multiple simultaneous spin flips at each annealing step as an analogue to quantum spin coherence as well as modifications of the Boltzmann acceptance probability to mimic quantum tunneling. We find that the use of multiple spin flips can indeed be advantageous under certain annealing schedules, but only for long anneal times.

## 1 Introduction

Simulated annealing is a classical numerical optimization technique based on the physical process of annealing used to slowly cool a material to its lowest energy state [1]. The method was first described by Metropolis and colleagues in 1953 [2] and later generalized by Hastings in 1970 [3] into what is now known as the Metropolis–Hastings algorithm, a subclass of more general Markov chain Monte Carlo methods [4].

A related optimization procedure, quantum annealing, has also been proposed for solving hard optimization problems [5,6]. In quantum annealing, the temperature is held fixed while the cost function slowly evolves from one with a simple structure to one representing the problem cost function. By the adiabatic theorem, if this process is performed slowly enough the system will eventually settle into a global minimum energy state [7,8]. Several generations of devices that implement quantum annealing for the Ising model have been built by D-Wave Systems, Inc. and used to solve a variety of optimization problems [9–13]. Whether such a device provides a quantum speedup over classical methods, and the mechanisms by which such a speedup would be possible, remain as yet unclear [14–16].

---

✉ Brian R. La Cour
   blacour@arlut.utexas.edu

[1] Applied Research Laboratories, The University of Texas at Austin, P.O. Box 8029, Austin, TX 78713-8029, USA

It has been suggested that the mechanism of tunneling may provide a speedup for quantum annealing over classical simulated annealing for coupled spin systems, as the coherent tunneling of multiple spin states would allow the system to escape deep local minima and explore the state space more widely [17–19]. Colloquially, tunneling allows one to pass *through* a barrier rather than having to jump *over* it. For a typical simulated annealing algorithm, each candidate state would be selected by choosing one spin at random to flip. Such modest steps would seem prudent in traversing a well-structured cost function but may prove insufficient when dealing with a virtually structureless landscape, as is typical of most hard optimization problems. This suggests that flipping several spins at once, in analogy with coherent quantum tunneling, may prove advantageous. On the other hand, such wild excursions could be suboptimal, or perhaps even disastrous.

This idea is not new. Based on insights from percolation theory, Swendsen and Wang suggested that lattice spin simulations may be performed more efficiently when large clusters are changed in a single move [20,21]. This approach was shown to work well when clusters are chosen based on physically motivated problems, such as ferromagnetic spin lattices, but proved ineffective for more general spin glass problems. Houdayer has developed a cluster algorithm for two-dimensional spin glasses that moves between isoenergetic clusters in order to explore the state space more widely while avoiding costly random rejection steps in the Metropolis–Hastings algorithm [22]. Based on this work, Zhu, Ochoa, and Katzgraber have developed a more general isoenergetic cluster algorithm that is efficient for any spatial dimension, including the Chimera graph structure of the D-Wave system [23]. All of these approaches leverage either the specific graph structure or additional computation to select clusters.

The use of multiple spin flips is similar to the $k$-opt strategy used in local optimization, wherein a neighborhood of size $k$ is examined to select the lowest energy state among $k$ neighbors of the current state [24]. When $k$ is determined adaptively, this heuristic often provides high-quality solutions in polynomial time, but some problem instances may require exponentially many iterations, as large neighborhoods must be searched exhaustively [25]. In Ref. [19] the $k$-opt procedure for fixed $k$ is referred to as "algorithmic tunneling", due to its similarity with quantum tunneling, and found to give improved performance over the popular differencing method of Karmarkar and Karp when applied to the number partitioning problem [26]. This suggests that large-neighborhood searches may indeed be advantageous if the cost of searching each neighborhood can be restricted.

In this paper, we consider a variation of the $k$-opt strategy in which the neighborhood is selected randomly, and may be as large as the state space, but is used only to select a single candidate state. We then examine the benefits, and tradeoffs, of such an approach as compared to a baseline simulated annealing procedure using only single-spin flips and a fast annealing schedule. Isakov and colleagues have considered a similar approach, wherein multiple spins flips are performed simultaneously and probabilistically based on the energy differences of individual spins [27]. Here we consider a variation of this approach wherein the candidate spin configuration is selected at random, independent of the energy cost. We also consider a modification of the standard Boltzmann acceptance probability to one resembling quantum tunneling probabilities and examine the relative impact of this modification.

The outline of the paper is as follows. Section 2 describes the baseline and modified simulated annealing algorithms considered. Their relative performance over a number of different problems and for large anneal times is considered in Sect. 3. In Sect. 4 we consider the relative impact on performance of reducing the anneal time, and we summarize our conclusions in Sect. 5.

## 2 Algorithm Description

The basic Metropolis algorithm may be described as follows. Given an energy cost function that we wish to minimize and an initial point in the corresponding state space, we select (by some means) a new candidate state. If the energy decreases, then the candidate state is accepted and the iterative process continues. Otherwise, the candidate state is accepted with a probability that decreases with a notional temperature parameter. In the original Metropolis algorithm, this probability is given by the Boltzmann factor. The random acceptance step allows the algorithm to escape local minima, but with a probability that decreases with temperature. Thus, if the system cools slowly enough, it will eventually settle into a global minimum [28].

In this paper, our state space will be defined by the set $\mathcal{S} = \{-1, +1\}^n$, notionally representing a system of $n$ spin-$\frac{1}{2}$ particles with $N = 2^n$ possible spin configurations. The cost function is based on the general Ising model, which, for $s \in \mathcal{S}$, takes the form

$$E(s) = -\left( \sum_{i=1}^{n} h_i s_i + \sum_{i=1}^{n} \sum_{j=1}^{n} J_{ij} s_i s_j \right), \tag{1}$$

where $h_i \in \mathbb{R}$ represents an external magnetic field applied to spin $i$ and $J_{ij} \in \mathbb{R}$ represents the coupling between spins $i$ and $j$. Thus, when $h_i$ and $s_i$ are of the same sign, the energy is lowered. Similarly, $J_{ij} > 0$ represents ferromagnetic coupling, which energetically favors similarly aligned spins, while $J_{ij} < 0$ represents antiferromagnetic coupling, which favors oppositely aligned spins. For the problems we shall consider, the coupling is always symmetric (i.e., $J_{ij} = J_{ji}$), though no special graph connectivity constraints are assumed. The optimization problem consists of finding at least one point $s_* \in \mathcal{S}$ such that $E(s_*) \leq E(s)$ for all $s \in \mathcal{S}$. The class of general Ising optimization problems, with no special coupling constraints, is known to be a nondeterministic polynomial (NP) hard [29].

By transforming $s$ to $x \in \{0, 1\}^n$, where $x_i = (s_i + 1)/2$, the Ising problem may be converted to an equivalent Quadratic Unconstrained Binary Optimization (QUBO) problem. We may then take $x$ to represent the binary digits of the integer $x \in \{0, \ldots, N-1\}$, where $x = x_1 2^{n-1} + \cdots + x_n 2^0$. In terms of $x$, the cost function takes a similar form to that of the Ising model.

The baseline simulated annealing (SA) algorithm we use may be described as follows. A fast annealing schedule is adopted such that, at time step $k \in \{1, \ldots, K\}$, the temperature is given by $T(k) = T_0/k$. The initial temperature, $T_0$, is taken to be the maximum energy, thereby allowing for large jumps at the beginning of the annealing process [30]. Thus, for $|J_{ij}| = 0$ the initial temperature, $T_0$, scales linearly with $n$, while for $h_i = 0$ it scales quadratically. Specifically, we use the upper bound

$$T_0 = \sum_{i=1}^{n} |h_i| + \sum_{i=1}^{n} \sum_{j=1}^{n} |J_{ij}| \geq E(s) . \tag{2}$$

Given a state $s(k) \in \mathcal{S}$ at time step $k$, we select a new, candidate state $s'(k) \in \mathcal{S}$ such that one of the $n$ spins, chosen at random, is flipped in sign. If $E(s'(k)) < E(s(k))$, this candidate state is accepted and $s(k + 1) = s'(k)$; otherwise, it is accepted with probability $p(s'(k), s(k), T(k))$, where

$$p(s', s, T) = \exp\left\{-\left[E(s') - E(s)\right]/T\right\} \ . \tag{3}$$

For $T = 0$, this probability is taken to be 1.

The modified algorithm, which we call simulated annealing with multiple simultaneous spin flips (SAM), is identical to the baseline SA algorithm described above with the exception that, when selecting a candidate state, a random number of spins (drawn uniformly from 1 to $n$) are chosen to be flipped. This allows for wider excursions in the state space, similar to what is done in fast annealing methods for continuous state spaces [31,32].

## 3 Performance Analysis

In what follows, we shall consider the performance of the baseline SA and proposed SAM algorithms relative to a number of different representative problems. First, a random realization of the problem is generated. For each of the two algorithms, the annealing procedure is then repeated a certain number of times, with a different, randomly selected initial state each time. The algorithms are applied to the same problem realization, and the resulting probability of success (i.e., of settling on one of the global minima) is computed. Upper and lower 95 % confidence bounds are computed using the (exact) Clopper-Pearson method [33].

The number of annealing steps is taken to be $K = N$, which is the minimum number required to traverse the entire state space at least once using only single-spin flips. This guarantees that a solution can always be found, even if the algorithm fails to find it.

### 3.1 False Minimum Problem

This problem, also known as the "weak-strong cluster networks" problem, arises from the work by Boixo, Denchev, and colleagues, who use it to study the effects of collective tunneling on quantum annealing [17,19]. The energy function has a global minimum at $x = 0$ (all spin down) and a false minimum at $x = N - 1$ (all spin up). The local magnetic fields are defined such that $h_i = 1 - \varepsilon > 0$ for $i \leq n/2$ and $h_i = -1$ for $i > n/2$. The couplings are defined as in Ref. [17] and correspond to a Chimera graph structure, thereby restricting $n$ to be a multiple of 4. The resulting energy gap between the false and true minima is $\Delta E = n\varepsilon$.

We considered examples of the false minimum problem for $\varepsilon = 0.1$ and $n \in \{4, 8, 12, 16\}$. Each of the four problem instances was repeated $R$ times: for $n = 4$, $R = 10\,000$; for $n \in \{8, 12\}$, $R = 1000$; for $n = 16$, $R = 100$. The results are shown in Fig. 1. As can be seen, the SA algorithm performs much more poorly than the SAM algorithm due to its propensity to get trapped in the false minimum. Interestingly, the SA performance becomes consecutively worse with increasing $n$, while the SAM algorithm's performance tends to increase with increasing problem size. This behavior is believed to be due to the fact that the fast annealing schedule used tends to limit the rate of acceptance and, hence, "freeze" the system early in the annealing process.

### 3.2 Zero Coupling Problem

It may be suspected that an overly aggressive move strategy may perform more poorly for "easy" problems in which a single spin flip can bring the state closer to the global minimum. We therefore compare the SA and SAM algorithms for the case in which there is no coupling between the spins (i.e., $J_{ij} = 0$ for all $i$, $j$), and the local magnetic fields $h_i$ are taken to $\pm 1$ with equal probability.

**Fig. 1** Plot of the probability of successfully finding the global minimum versus the number of spins for the false minimum problem

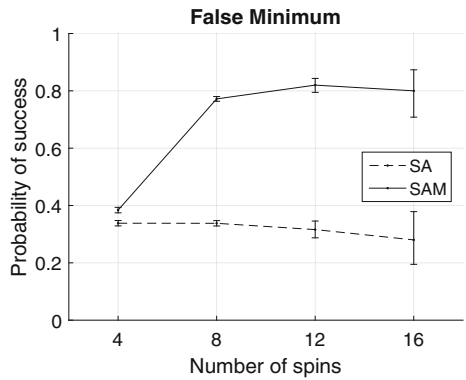

**Fig. 2** Plot of the probability of successfully finding the global minimum versus the number of spins for the zero coupling problem

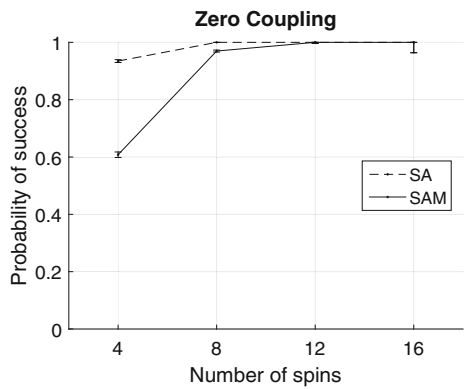

We again consider examples for $n \in \{4, 8, 12, 16\}$. For each value of $n$, a single problem realization is drawn and used for both algorithms. The results are shown in Fig. 2. As expected, the SA algorithm performs better than the SAM algorithm, particularly for low values of $n$, but the two quickly converge in performance for $n$ greater than about 12.

### 3.3 Uniform Spin-Glass Problem

The spin-glass problem takes its inspiration from a physical spin glass consisting of a collection of disordered spins, itself analogous to the disordered positions of a glassy material [34,35]. The uniform spin-glass problem is characterized by having $h_i = 0$ for all $i$ and $J_{ij} = \pm 1$ with equal probability. By convention, $J_{ii} = 0$ and $J_{ji} = J_{ij}$. This problem is an example of a frustrated spin system and has been studied extensively in the context of both simulated and quantum annealing [12]. We consider spin-glass problems on a fully connected graph (i.e., one in which every spin is connected to every other spin), as more restricted graph structures may be solved more easily [36,37].

The results of a single realization for each of the four values of $n$ is shown in Fig. 3. There we see that the performance of SA is somewhat better than that of SAM for low values of $n$ (around 4), but for larger values the opposite is true. In particular, the probability of success for SA tends to decrease with increasing $n$, while that of SAM tends to increase, saturating to near unity for *n* greater than about 12. Although we show a single problem instance here for illustrative purposes, this behavior was found to be typical of random problem instances, as described later in Sect. 3.5.

**Fig. 3** Plot of the probability of successfully finding the global minimum versus the number of spins for the uniform spin-glass problem

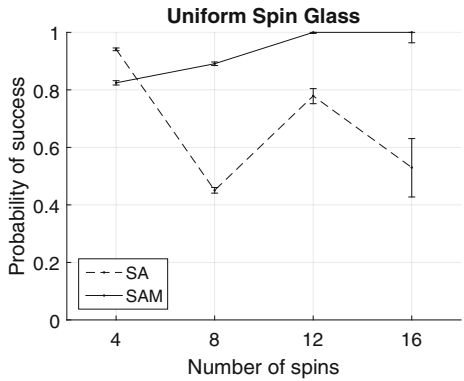

### 3.4 Gaussian Spin-Glass Problem

The final type of problem considered was a Gaussian version of the spin-glass problem. The Gaussian spin-glass problem is similar to the uniform spin-glass problem, except that the coupling $J_{ij}$ is now taken to be Gaussian distributed with zero mean and unit standard deviation. Because of the continuous distribution of coupling strengths, the gap between neighboring energy levels can be arbitrarily small, and, as *n* increases, this gap will tend to decrease. Consequently, we expect that the global minimum will be more difficult to find than in the case of the uniform spin-glass problem.

The results, summarized in Fig. 4, tend to follow this general expectation. As in the uniform spin-glass problem considered previously, the performance of SA is slightly better than that of SAM for small values of *n* (about 4), while SAM tends to outperform SA for larger problems. However, the performance of both methods tends to decrease with increasing *n* for values larger than about 8.

**Fig. 4** Plot of the probability of successfully finding the global minimum versus the number of spins for the Gaussian spin-glass problem

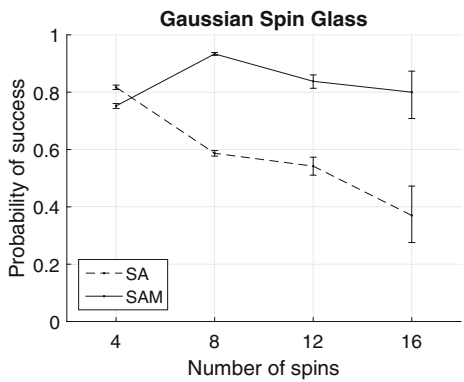

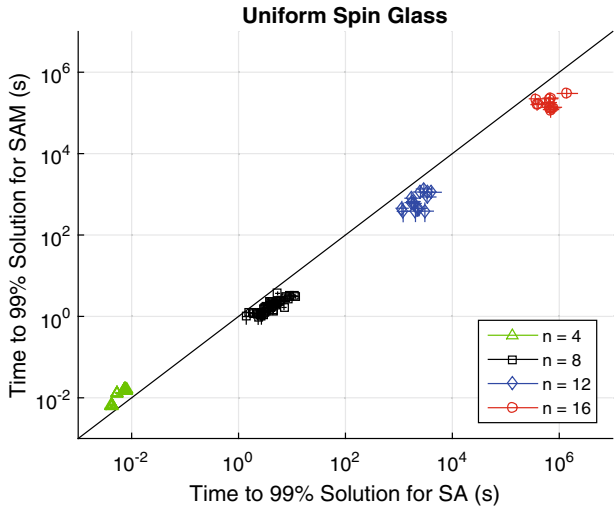

**Fig. 5** Plot of the time to solution for the SAM versus SA methods for randomized instances of the uniform spin-glass problem. The *symbols* and *colors* indicate different problem sizes: $n = 4$ (*green triangles*), $n = 8$ (*black squares*), $n = 12$ (*blue diamonds*), $n = 16$ (*red circles*). The *horizontal* and *vertical lines* indicate the 95 % confidence intervals (Color figure online)

### 3.5 Randomized Problems

The previous subsections considered a single problem instance for each value of $n$ and compared the probability of success for the two methods SA and SAM as a function of problem size for multiple repetitions of the same problem instance. There we found that, typically, SAM performs better than SA, at least for the more difficult problems (i.e., $n > 4$). One may wonder how typical these results are over an ensemble of problem instances.

To address this question we considered 100 random realizations of the two spin-glass problems. For each problem realization, the two methods SA and SAM were run over multiple repetitions, as before. To compare them, we plot the time to solution for a 99 % probability of success for SAM versus that of SA, with a diagonal line indicating equivalency of the two methods. Given a probability of success $p_s$ and total annealing time $t_a$, the time to solution is given by $t_a \log(1 - 0.99)/\log(1 - p_s)$, which represents the average number of repeated anneals, and their cumulative time, to achieve an overall 99 % probability of success.

The results for the uniform spin-glass problem are shown in Fig. 5, where we can see that most of the points, and all those for which $n > 4$, fall below the diagonal line, indicating that SAM performed better than SA. Similar results are found for the Gaussian spin-glass problem, as shown in Fig. 6. In this case, there is a good deal more variation across problem realizations, but there is a clear advantage of SAM over SA for $n > 4$.

### 3.6 Quantum-like Tunneling

Finally, we considered the effect of replacing the Boltzmann transition probability of Eq. (3) with one similar to a quantum barrier penetration probability, given by

$$p_Q(s', s, T) = \exp\left\{-d(s', s)\sqrt{[E(s') - E(s)]/T}\right\}, \tag{4}$$

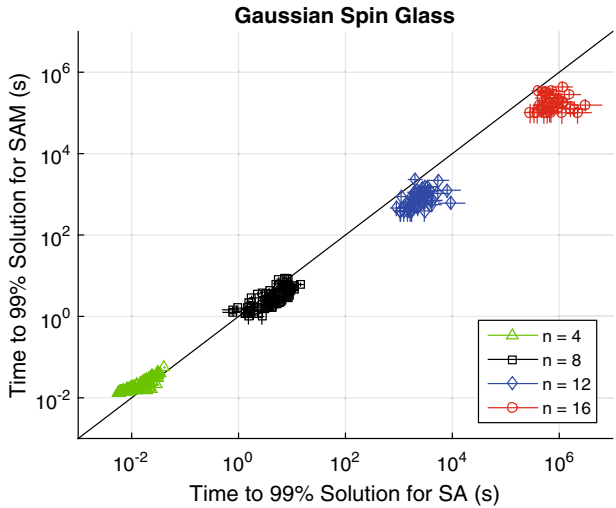

**Fig. 6** Plot of the time to solution for the SAM versus SA methods for randomized instances of the Gaussian spin-glass problem. The *symbols* and *colors* indicate different problem sizes: $n = 4$ (*green triangles*), $n = 8$ (*black squares*), $n = 12$ (*blue diamonds*), $n = 16$ (*red circles*). The *horizontal* and *vertical lines* indicate the 95 % confidence intervals (Color figure online)

where $d(s', s)$ is the Hamming distance between spin states $s'$ and $s$ [38]. This probability captures the quantum dependence of tunneling through a region with a "barrier width" of $d(s', s)$ versus jumping over a barrier of height $E(s') - E(s)$. It also captures the dependence on the square root of the energy difference vice the absolute difference, thereby effectively lowering the barrier height through a nonlinear transformation.

We considered the randomized problem instances described previously, using the candidate state selection as for the SAM method but replacing the Boltzmann transition probability with that of Eq. (4), and call this method SAQ (Simulated Annealing with Quantum-like tunneling). The results are shown in Figs. 7 and 8 for the uniform and Gaussian spin-glass problems, respectively. In both cases, we find nearly equal performance, indicating that the choice of Boltzmann versus quantum-like transition probabilities appears to make little difference, especially for the harder problems. Thus, the dominant mechanism in both methods appears to be coherent spin flips rather than tunneling-like behavior.

## 4 Reduced Annealing Times

The comparisons of the previous section used a large number of annealing steps, $K = N$. With this many steps, one could of course simply evaluate the cost function at every possible value and thereby be guaranteed a solution. Practically, a much smaller number of steps will typically be used. If we select $K \leq N$ unique spin configurations at which to evaluate the cost function, then, since there are two possible solutions (when $h_i = 0$), the probability of obtaining at least one of them using this brute-force (BF) approach is

$$p_K = \frac{2\binom{N-2}{K-1} + \binom{N-2}{K-2}}{\binom{N}{K}} = \frac{K(2N - K - 1)}{N(N - 1)}.$$ (5)

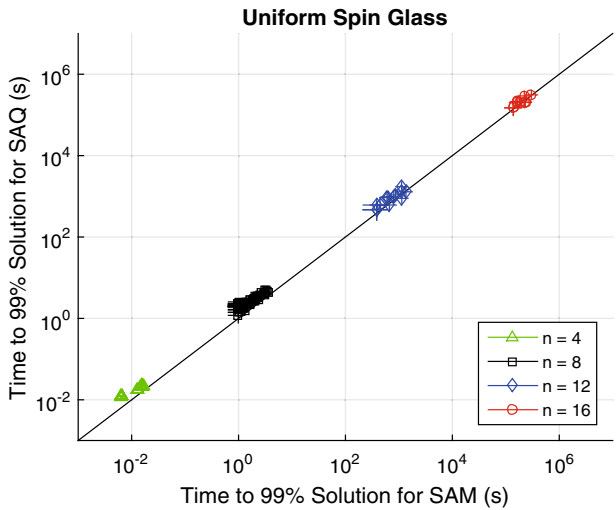

**Fig. 7** Plot of the time to solution for the SAQ versus SAM methods for randomized instances of the uniform spin-glass problem. The *symbols* and *colors* indicate different problem sizes: $n = 4$ (*green triangles*), $n = 8$ (*black squares*), $n = 12$ (*blue diamonds*), $n = 16$ (*red circles*). The *horizontal* and *vertical* indicate the 95 % confidence intervals (Color figure online)

**Fig. 8** Plot of the time to solution for the SAQ versus SAM methods for randomized instances of the Gaussian spin-glass problem. The *symbols* and *colors* indicate different problem sizes: $n = 4$ (*green triangles*), $n = 8$ (*black squares*), $n = 12$ (*blue diamonds*), $n = 16$ (*red circles*). The *horizontal* and *vertical lines* indicate the 95 % confidence intervals (Color figure online)

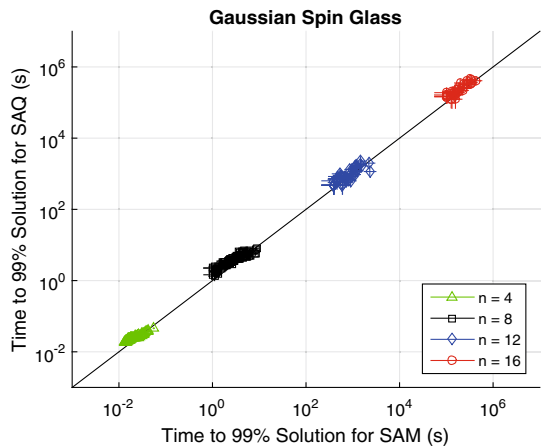

Note that $p_N = 1$ for $K = N$ and $p_1 = 2/N$ for $K = 1$, as expected.

So, how does this compare to the probabilities of success for the SA, SAM, and SAQ methods? We examined this question by computing the probability of success over 100 realizations of the Gaussian spin-glass problem, much as was done in the previous section, but with lower values of $K$. Figure 9 plots the probabilities of success for SA, SAM, SAQ, and BF versus the relative annealing time $K/N$ for two values of $n$ using random instances of the Gaussian spin-glass problem. From these plots, several observations can be made. For $K/N$ near unity, the brute-force approach is, of course, the best of the three, although a more optimal annealing schedule would give nearly the equal performance. As $n$ increases, though, the SAM and SAQ methods quickly approach ideal performance, while the SA method experiences decreasing performance, much as was seen in the previous examples.

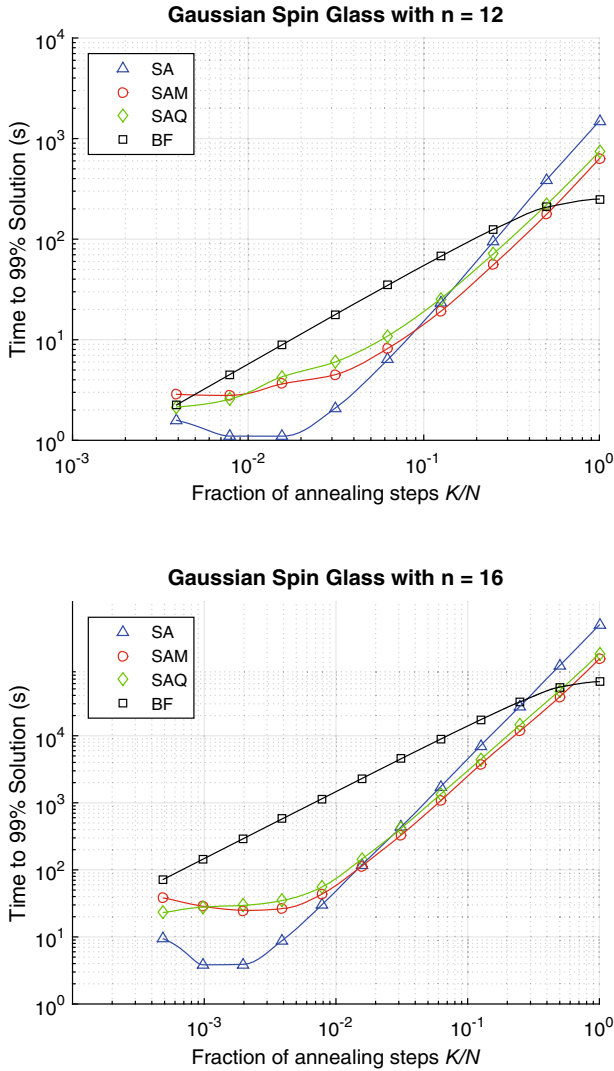

**Fig. 9** Plot of the time to solution for the Gaussian spin-glass problem versus the relative annealing time $K/N$ for $n = 12$ spins (*top plot*) and $n = 16$ spins (*bottom plot*). The *three curves* correspond to spline fits for the the SA (*triangles*), SAM (*circles*), and BF (*squares*) methods (Color figure online)

There is, however, a crossover point below which the brute-force approach is no longer advantageous. This point increases with $n$ for SAM and SAQ but decreases with $n$ for SA. Thus, for near-maximal annealing times ($K \lesssim N$), BF is slightly better than SAM and SAQ, which is much better than SA.

For intermediate annealing times, SAM and SAQ are superior to SA, and both are superior to BF. There is, however, another crossover point such that, for any smaller number of annealing steps, SA outperforms SAM and SAQ. In this regime, SA benefits from locally optimal searches that, over many repetitions, start at different points in the configuration state space. SAM and SAQ, by contrast, search too widely for the limited annealing time allotted and, therefore, have trouble finding an optimal solution, even if they start near one. We note that the crossover point at which SA, SAM, and SAQ are equal corresponds to a relative annealing time $K/N$ that decreases with increasing problem size, even though the total number of annealing steps, $K$, continues to increase. For the Gausian spin-glass problem, this crossover point is about $K/N \approx 8\%$ for $n = 12$ and $K/N \approx 1\%$ for $n = 16$.

The above examples considered success on a single instance of running the given algorithm for a certain annealing time $K$. One may also consider running the algorithm multiple times using a shorter annealing time $K' \leq K$ such that $K/K' \in \mathbb{N}$ and taking the best result (i.e., the configuration giving the lowest energy). For the brute-force approach, it can be shown that this strategy is never advantageous; in other words, for all $K'$ we have $1 - (1 - p_{K'})^{K/K'} \leq p_K$, since $p_K$ drops sharply with decreasing $K$. The probabilities of success for the SA, SAM, and SAQ methods initially drop much more slowly with decreasing $K$ and, so, for relatively large values of $K'$ it may be quite advantageous to perform multiple shorter anneals. For very small values of $K'$, however, the probability of success decreases more rapidly with decreasing $K'$, eventually devolving to the brute-force approach, and, so, this becomes no longer an advantageous strategy. Thus, performing multiple repetitions of the simulated annealing process, for either method, can prove to be of great benefit, but care must be taken that the probability of success on any one instance is not too low (i.e., close to $p_K$).

## 5 Conclusions

In this paper we have considered and compared three different approaches (SA, SAM, SAQ) to performing classical simulated annealing on discrete spin systems. The baseline SA approach considers only one spin flip per annealing step, while the proposed SAM and SAQ methods flip a random and unbounded number of spins at each step. Both SAM and SAQ use an analogue of quantum spin coherence, with the former using a Boltzmann acceptance probability and the latter using an analogue to the quantum tunneling probability. Performance was considered using a variety of problem types and instances based on the general Ising model with arbitrary spin coupling, which were then solved numerically.

Based on this analysis, we find that the SAM and SAQ methods generally outperform the SA method when the problem is difficult and the number of annealing steps (i.e., the annealing time) is large. For easier problems or smaller annealing times, SA typically performs better than both SAM and SAQ. This may be interpreted to mean that when there is little time to search it is usually best to search locally. Of course, SA and SAM are two extremes in a spectrum of approaches, and an optimal strategy would likely use a number of simultaneous spin flips that increases with the anneal time, together with an optimized annealing schedule. The SAM and SAQ methods themselves perform nearly identically, suggesting that multiple simultaneous spin flips are the dominant mechanism for their performance.

We also found that it can be advantageous to perform multiple, shorter anneals rather than a single anneal with the same total time. For very short anneal times, however, the probability of success approaches that of a simple brute-force search and, hence, the advantage of multiple short anneals becomes lost. Thus, care must be taken in using an annealing procedure that individually gives a very low probability of success.

Finally, we note that, to the extent that these analogues do mimic behavior in a true quantum annealing system, a quantum speedup, relative to the baseline SA method, will only be realized for large anneal times. Of course, if the coherence times are much shorter than the anneal times, this advantage may never be realized. Nevertheless, using multiple spin flips it may be possible to mimic such a speedup by means of a suitable classical system.

**Acknowledgments** This work was supported by the Office of the Secretary of Defense under Contract No. N00024-07-D-6200-0743 and by Applied Research Laboratories, The University of Texas at Austin, under an Independent Research and Development Grant.